\documentclass[twocolumn,times,tighten]{aastex63}
\usepackage{color}
\usepackage{enumitem}
\usepackage{hyperref}
\usepackage{verbatim}
\usepackage{amsmath,amstext,mathrsfs}
\usepackage[all]{hypcap} 
\usepackage{afterpage}    
\usepackage{marginnote}
\usepackage{makecell}
\usepackage{subfigure}

\usepackage{wrapfig}

\newcommand*{\toreferee}{} 
\graphicspath{{./}{figures/}}

\received{...}
\revised{...}
\accepted{...}
\submitjournal{ApJ}

\shorttitle{Fast modes and gradients}
\shortauthors{Ho \& Lazarian}

\graphicspath{{./}{figures/}}

\begin{document}

\title{Intermittency of fast MHD modes and regions of anomalous gradient orientation in low $\beta$ plasmas
\footnote{Drafted on Aug, 31st, 2020}}
\email{kho33@wisc.edu, lazarian@astro.wisc.edu}

\author{Ka Wai Ho}
\affiliation{Department of Astronomy, University of Wisconsin-Madison, USA}

\author{A. Lazarian}
\affiliation{Department of Astronomy, University of Wisconsin-Madison, USA}
\begin{abstract}
The high alignment of small-scale turbulent Alfv\'{e}nic motions with the direction of magnetic field that percolates the small scale eddies   imprint the direction of the magnetic field are the properties that follow from the MHD theory and the theory of turbulent reconnection. The Alfv\'{e}nic eddies mix magnetic fields perpendicular to the direction of the local magnetic field and this type of motions is used to trace magnetic fields with the Velocity Gradient Technique (VGT). The other type of turbulent motions, fast modes, induces anisotropies orthogonal to Alfv\'{e}nic eddies and interferes with the magnetic field tracing with the VGT. We report a new effect, i.e. in magnetically dominated low-$\beta$ sub-sonic media, fast modes are very intermittent and that in a volume with a small filling factor the fast modes dominate other turbulent motions. We identify these localized regions as the cause of the occasional change of the direction of gradients in our synthetic observations. We show that the new technique of measuring the Gradients of Gradient Amplitudes (GGA) suppresses the contribution from the fast mode dominated regions, improving the magnetic field tracing. In addition, we show that the distortion of the gradient measurements by fast modes is also applicable to the synchrotron intensity gradients (SIGs), but the effect is reduced compared to the VGT. 
\end{abstract}
\keywords{Interstellar magnetic fields (845); Interstellar medium (847); Interstellar dynamics (839);}

\section{Introduction} \label{sec:intro}

Turbulence is ubiquitous in astrophysical environment. Magneto-hydrodynamics (MHD) turbulence plays a crucial role in various astrophysical phenomena, including the formation of stars (see \citealt{MO07,MK04}), the propagation and acceleration of cosmic rays (see \citealt{J66,Sch10,YL08}), remove angular
momentum from accretion disks (see \citealt{Krasnopolsky12} ),the regulation of heat and mass transfer between different ISM phases (see \citealt{D09} for the list of the different ISM phases). 

Over the decades, the progress of the theory improved our understanding of MHD turbulence \citep{GS95,LV99} and numerical studies Cho \& Vishniac 2000, Maron \& Goldreich 2001, \citep{CL02,CL03} help to test the proposed theoretical ideas. A detailed discussion of the present understanding of MHD turbulence can be found in \cite{BL19}. Based on the advancement of MHD theories, techniques such as correlation function analysis (CFA) and the principal component analysis of anisotropies (PCAA, \cite{2008ApJ...680..420H}) were proposed to study magnetic field statistically using the theoretical understanding of anisotropic scaling \citep{L02,EL05,2008ApJ...680..420H,EL11,Burkhart14}. They showed the ability to trace the magnetic field using anisotropy in the MHD simulations and observation.

The most promising way of magentic field tracing is related to a recently proposed technique, velocity gradient technique (VGT), is also capable of tracing the magnetic field. The technique also makes use of the fact that magnetic field makes turbulence anisotropic, with turbulent eddies elongated along the magnetic field (See \cite{BL13} for a review). As a result, it induces the fluid motion perpendicular to the local magnetic field direction and makes the gradients of velocity become perpendicular to the local direction of the magnetic field. This property of magnetic turbulence was employed in \cite{GL17}, in which the approach using velocity centroid gradients (VCGs) to trace magnetic field orientations was proposed. With the sub-block averaging approach proposed in \cite{YL17a}, the technique showed its high potential. 

The VGT has been numerically tested for a wide range of column densities from diffuse transparent gas to molecular self-absorbing dense gas.  It was shown to be able to provide both the orientations of the magnetic field as well as a measure of media magnetization (see \cite{LY18a,LYH18}). A VGT survey \citep{HuNature} was conducted recently to study the morphology of five low mass molecular cloud. The result is consistent with the Planck polarization measurement and successfully showed the reliability of the VGT technique. 

Other than tracing magnetic field orientation and the magnetization, gradient technique can also trace the signature of important astrophysics process imprint in the fluid, for instance, in the presence of gravity and shocks  (see \cite{YL17,Hu2020}). The observational signature of this change in the cloud dynamics is that the VGT orientation flips by up to 90 degrees to align parallelly with the direction of the magnetic fields. We call this term as  orthogonal gradient. This open a novel way for the community to study the strong region through VGT. However, similar signature of orthogonal gradient could also happen due to the  compressible nature of MHD turbulence (see \cite{CL03,LY18a}), in particular due to the fast modes.  

In this paper, we would like to explore the relation between fundamental properties of the MHD turbulence and orthogonal gradient that appear in gradient technique. In what follows, we would cover the theory in section \ref{sec:theory} and our numerical setup in section \ref{sec:method}. Then we would discuss the relation between MHD modes and properties of gradient in sections \ref{sec:Mode properties} and \ref{sec:gradient modes}. We further study the impact of fast modes to both 3D and 2D observable measures in section \ref{sec:Intermittency}. We than introduce the new techniques to reduce the impact of fast modes in the 2D observable measures in section \ref{sec: tech}. At last, we would discuss our work in section \ref{sec:discussion} and summarize the paper in section \ref{sec:summary}.

\section{Properties of MHD turbulence }
\label{sec:theory}
MHD turbulence theory is the subject which development has been boosted due to the ability of preforming high resolution numerical simulations. The original studies of Alfv\'{e}nic turbulence in  \cite{Iron1964} and \cite{Kraichnan1965} were based on a hypothetical model isotropic MHD turbulence. The numerical studies, however, demonstrated the anisotropic nature of the MHD cascade. 

\subsection{Anisotropy in incompressible MHD turbulence}
The predictive modern theory of incompressible MHD turbulnce was formulated in \citealt{GS95} (hereafter GS95). While the theory initially gained lukewarm response by the MHD turbulence community,  later theoretical and numerical studies extended the studies and provided rigorous testing. In this process the theory was augmented with the essential concept of {\it local} direction of magnetic field. This concept most is most naturally derived from the properties of magnetic eddies undergoing fast turbulent reconnection \cite{LV99} (hereafter LV99). Our present study is based on the modern understanding of the MHD turbulence cascade, and the statistical properties of MHD turbulence that are confirmed numerically. 

In the MHD theory, the Alvenic mode cascade exibits its anisotropic nature at the sub-Alfv\'{e}nic regime, i.e. for the injection velocity $V_L$ being less than the Alfvén velocity $V_A$. Namely, the Alfvén modes initially evolve along the so-called weak turbulent cascade, i.e. increasing the perpendicular wave-number $k_{\bot}$ while keeping the parallel wave-number $k_{\parallel}$ the same (see LV99, \cite{Galtier2005}). However, at sufficiently small scales the nature of the cascade changes. The magnetic eddies that aligned with the local magnetic field surrounding the eddies, mix up this field and out Alfv\'{e}nic waves with a period equal to the period of an eddy, which is:

\begin{equation}
\label{eq:Alfenc_turning}
\begin{aligned}
l_{\bot}/v_{l} \approx l_{\parallel}/V_{A}
\end{aligned}
\end{equation}
where $l_{\parallel}$ is the parallel scale of the eddy and $l_{\bot}$ is the eddy size perpendicular to the magnetic field. In GS95, this condition comes with the critical balance, indicate the fact of assuming incompressible nature, meaning zero velocity divergence throughout the entire space. As a result, the infall velocity gradient $v_{l}/l_{\bot}$ should be equivalent to that of the propagating velocity gradients $V_{A}/ l_{\parallel}$ of the Alfvénic wave along the magnetic field line. Combining with velocity scaling $v_{l} \sim l^{1/3}$, we could obtain the vital scaling  relation between the parallel and perpendicular scales of the eddies,

\begin{equation}
\label{eq:Alfenc_turning}
\begin{aligned}
l_{\parallel} \sim l_{\bot}^{2/3}
\end{aligned}
\end{equation}

The anisotopy scaling indicates that increase of the perpendicular wave-number  makes the Alfvénic wave vectors more and more perpendicular to magnetic field. The physical picture, is that the turbulent eddies are getting more and more elongated along the direction of the magnetic field as we zoom into smaller scales. 

At the same time, one can see that the velocity gradients following the same anisotropy scaling, meaning the largest gradients correspond to the smallest and most elongated eddies. Later studies \citep{Letal17} shows the properties of velocity gradient could also applied to the gradients of the turbulent magnetic field. So, any observable quantities containing the information of magnetic field and velocity could enable the tracing of local magnetic field direction.

\subsection{Compressible MHD turbulence}

In the case of compressible turbulence, three MHD modes arises, namely incompressible Alfv\'{e}n mode and the two
compressible modes called fast and slow modes. We use the word “modes” rather than “waves” as, in strong MHD turbulence, the properties of motions may not be wave like while Alfvénic modes are essentially eddies, and they non-linearly decay within one period. The first study of anisotropy of the three modes were proposed by \cite{CL02,CL03} through the decomposition method. The corresponding equations determining the basis for the decomposition into modes are
\begin{equation}
\label{eq:slow_mode_presentation}
\begin{aligned}
\hat{\xi}_A \propto \hat{k}_{\parallel} \times \hat{k}_{\bot}\\
\hat{\xi}_s \propto \Big(1+\frac{\beta}{2}-\sqrt{D}\Big)k_{\bot}\hat{k}_{\bot}+\Big(-1-\frac{\beta}{2}-\sqrt{D}\Big)k_{\parallel}\hat{k}_{\parallel} \\
\hat{\xi}_f \propto \Big(1+\frac{\beta}{2}+\sqrt{D}\Big)k_{\bot}\hat{k}_{\bot}+\Big(-1+\frac{\beta}{2}+\sqrt{D}\Big)k_{\parallel}\hat{k}_{\parallel} \\
\end{aligned}
\end{equation}
where $D = (1+\beta/2)^2-2\beta cos^2\theta$, $\beta = 2M_A/M_S$, $cos\theta = k_{\parallel}\cdot \hat{{\bf B}}$, and $\hat{\xi}$ is the displacement vector with the  subscript indicate its mode ($A$:  Alfv\'{e}n, $f$: Fast, and $s$: Slow). 

\cite{CL03} performed the numerical analysis for both gas-pressure-dominated (high $\beta$ regime) and magnetic-pressure-dominated (low $\beta$ regime) plasma. They showed for both regime,  Alfv\'{e}n and slow mode follows the GS95 scale-dependent anisotropy and the same Kolmogorov $E(k) \sim k^{5/3}$ cascade, fast mode exhibit $k^{3/2}$ spectrum and isotropic scaling. 

Apart from the scaling, \cite{CL03} also showed the cascade of Alfv\'{e}n modes is almost independent of the slow and fast modes. The amount of Alfv\'{e}nic mode's energy drained into compressible modes is negligible, which means they are having their own cascade and evolve independently. 

In addition, Alfvén and slow modes together carry most of the energy of the turbulent cascade. In this case, tracing the gradient of velocity/magnetic field of Alfv\'{e}nic and slow mode could provide the local direction of the magnetic field.  Fast modes, in many cases, are subdominant in terms of the energy cascade, although they play a very important role for a number of key astrophysical processes, e.g., cosmic ray scattering.( see \cite{YL02,YL04}). However, the study of fast mode mainly focus on the statistical study while their spatial property have not been {\toreferee discussed} or explored, such as spatial energy distribution. The latter is performed in the current study.


\section{Numerical simulations}
\label{sec:method}
We study the MHD modes and gradient using the numerical simulations obtained by two MHD codes. The first one is the 3D MHD compressible, single fluid, operator-split, staggered grid MHD Eulerian code ZEUS-MP/3D \citep{Hayes06} to set up three-dimensional, uniform turbulent medium. We use a range of Alfv\'{e}nic Mach number $M_A=V_L/V_A$ and sonic Mach number $M_s=V_L/V_s$, where $V_L$ is the injection velocity; $V_A$ and $V_s$ are the Alfv\'{e}n and sonic velocities respectively. The second one is the advanced MHD simulation code, Athena++.  Athena++ is a complete re-write of the Athena MHD code \citep{Stone08} in C++. Compare to Zeus, the algorithm of Athena++ is based on directionally-unsplit, higher-order Godunov methods, which not only are ideal for use with both SMR and AMR, but also are superior for shock capturing. We employ Zeus for normal low $\beta$ simulation and Athena++ for extreme low beta simulation. To study the property of mode in low-$\beta$ environment and prevent the density effect caused by the supersonic turbulence, we selected the sub-sonic environment. Both of the simulations are isothermal with operation time of two sound crossing time ($\tau_{cs}$) such that the turbulence is fully developed. {\toreferee The turbulence is injected solenoidally for all the simulations using the Fourier-space method. Turbulent energy is injected at the large scale ( k=2 ) and dissipate the energy through the energy cascade.} To study the properties of modes  during the development of turbulence, we also perform a simulation with similar numerical parameters of H0S using Athena++ but lower resolution. For this simulation, we dump out a snapshot every 0.1 $\tau_{cs}$  until 3 $\tau_{cs}$. The numerical parameters are listed in Table \ref{tt1} in sequence of ascending values of media magnetization $ \beta = 2 (M_A/M_S)^2 $. Table \ref{tab:simulationparameters} shows the detail setup and of the simulation we used in this paper. {\toreferee The physical condition of the simulation can be seen as the emulation of the turbulent media that is warm and highly magnetized, such as the warm phase HI media and the synchrotron electron gas.} For all of the simulations, the mean magnetic direction points toward the z-axis and we define the line of sight direction as x-axis.
\linebreak

\begin{table}
 \centering
 \label{tab:simulationparameters}
 \begin{tabular}{c c c c c }
Model & $M_s$ & $M_A$ & $\beta=2(\frac{M_A}{M_s})^2$ & Code\\ \hline \hline
H0S & 0.70 & 0.07 & 0.02  & Athena++\\
H0SS & 0.76 & 0.11 & 0.021  & Athena++\\
H1S & 0.64 & 0.38 & 0.71  & Zeus\\
\label{tt1}
\end{tabular}
\caption { Simulations used in this paper. Resolution of them are $792^3$ for H0S/H1S and $480^3$ for H0SS. The $M_S$ and $M_A$ is computed at the final snapshot for each simulation (t = 2.0 $\tau_{cs}$ for H0S and H1S, t = 3.0 $\tau_{cs}$ for H0SS). A mean B-field direction points toward the z-axis for all the simulations.}
\end{table}

\section{Anisotropy and Fast and Slow mode Intermittency}
\label{sec:Mode properties}
To illustrate the properties of the MHD modes, we perform the mode decomposition similar to that in \cite{CL03}. For computing the specific component $b_i$ of each modes, the modes can be calculated as :

\begin{equation}
\label{eq:modes_computation}
\begin{aligned}
b_{(f,s,a),i} =  \mathfrak{F}^{-1} (\mathfrak{F}(\bf{b})\cdot \hat{\xi}_{f,s,a})(\hat{\xi}_{f,s,a} \cdot \hat{\xi}_{i})),
\end{aligned}
\end{equation}
where $\mathfrak{F}$ is the Fourier transfer operator, $b$ is the compute quantifies like velocity or magnetic field in this case, and {\toreferee subscript $i$ represents the direction such that $i\in \{x,y,z\}$}. The upper panel of Figure \ref{fig:Mode_illustration} illustrates the decomposition procedure that takes place in Fourier space.

\subsection{Anisotropy of MHD modes}
Several studies have been done on the anisotropy of different MHD modes \citep{CL03,LY18a,LY18b}. Here, we summarize the specific anisotropy of each mode briefly.

\begin{figure*}[t]
\label{fig:Mode_illustration}
\centering
\includegraphics[width=0.64\paperheight]{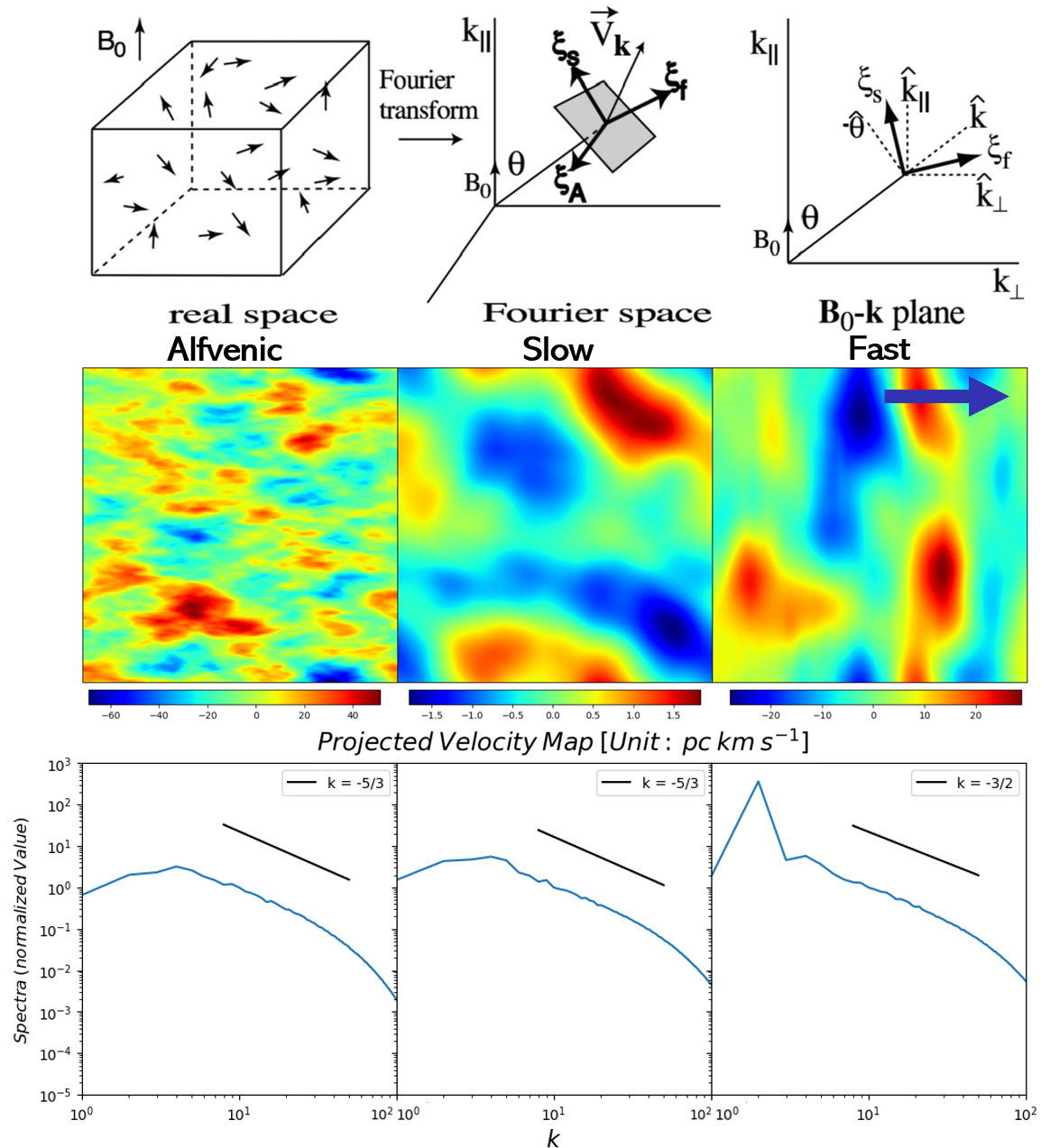}
\caption{Upper panel: Illustration of the decomposition method extracted from \cite{CL03}. The meaning of symbol stated in section \ref{sec:theory}. \\
Middle panel: The resultant projected maps of three velocity modes using the decomposition method from \cite{CL03} using the simulation cube H1S (with $M_s \approx 0.64$ and $M_A \approx 0.38$, $\beta \approx 0.7$ ). The blue arrow on the top right corner indicate the mean magnetic field direction for the simulation\\
Bottom panel: Energy spectra of each mode with black line marked the spectrum slope}
\end{figure*}

The middle panel of Figure \ref{fig:Mode_illustration} visualizes the anisotropy of three modes and the bottom panel shows their slope of the energy spectrum. We show the results of the decomposed velocity cube, which is projected along the x-axis (denoted as LOS-axis in this paper).

Alfvénic modes cascade on the scale of the order of one period, with the wave vector of the Alfvénic perturbations in strong turbulence being nearly perpendicular to the local direction of the magnetic field. As a result, the anisotropy and the iso-contours of intensity correlation are both elongated parallel to the magnetic field. This feature is shown in Figure \ref{fig:Mode_illustration}.

For slow modes, slow waves present perturbations that propagate along magnetic field lines. In the limit of incompressible media, slow waves are pure magnetic compression that propagate along magnetic field lines. Formally, the incompressible case corresponds to $\beta = \infty$, and in this limit, the slow modes are frequently called pseudo-Alfvén modes. By contrast, for $\beta \ll 1$, the slow waves are density perturbations propagating along magnetic field lines. Those anisotropy of perturbations would be imprint in the velocity as shown in  Figure \ref{fig:Mode_illustration}.

The properties of fast modes are rather  different for the high and low $\beta$ regimes. For high-$\beta$ case, the propagation of fast wave is similar to sound wave irrespective of its relation to the magnetic field. Whereas in the low-$\beta$ case, the propagation of fast wave corresponds to the magnetic field compression that propagate with Alfvén velocity. In this case, the anisotropy of fast mode would perpendicular to the magnetic field, which is also shown in Figure \ref{fig:Mode_illustration}. Therefore, we would also expect velocity eddies to be perpendicular to the region that is dominated by the fast modes, which is in reverse relation to that of Alfvénic turbulence.

\subsection{Intermittency of fast modes }
Numerical studies (\cite{CL02,CL03,LY18a,LY18b}) indicate that the fast modes are sub-dominat at least for the cases of incompressible driving of turbulence (e.g. \cite{LY18a}). For our simulations, we also found this property holds on the global scale. We noticed that most of the energy are distributed through the Alfvén modes and slow modes in the case of low-$\beta$ environment with the amount of about 50 \% for Alfvénic modes, 35\% for slow modes and only 15\% for fast modes. This shows that the fast mode play a subdominant role {\it on average}. 

\begin{figure}
\label{fig: Fast_Intermittency_2D}
\includegraphics[width=0.49\textwidth]{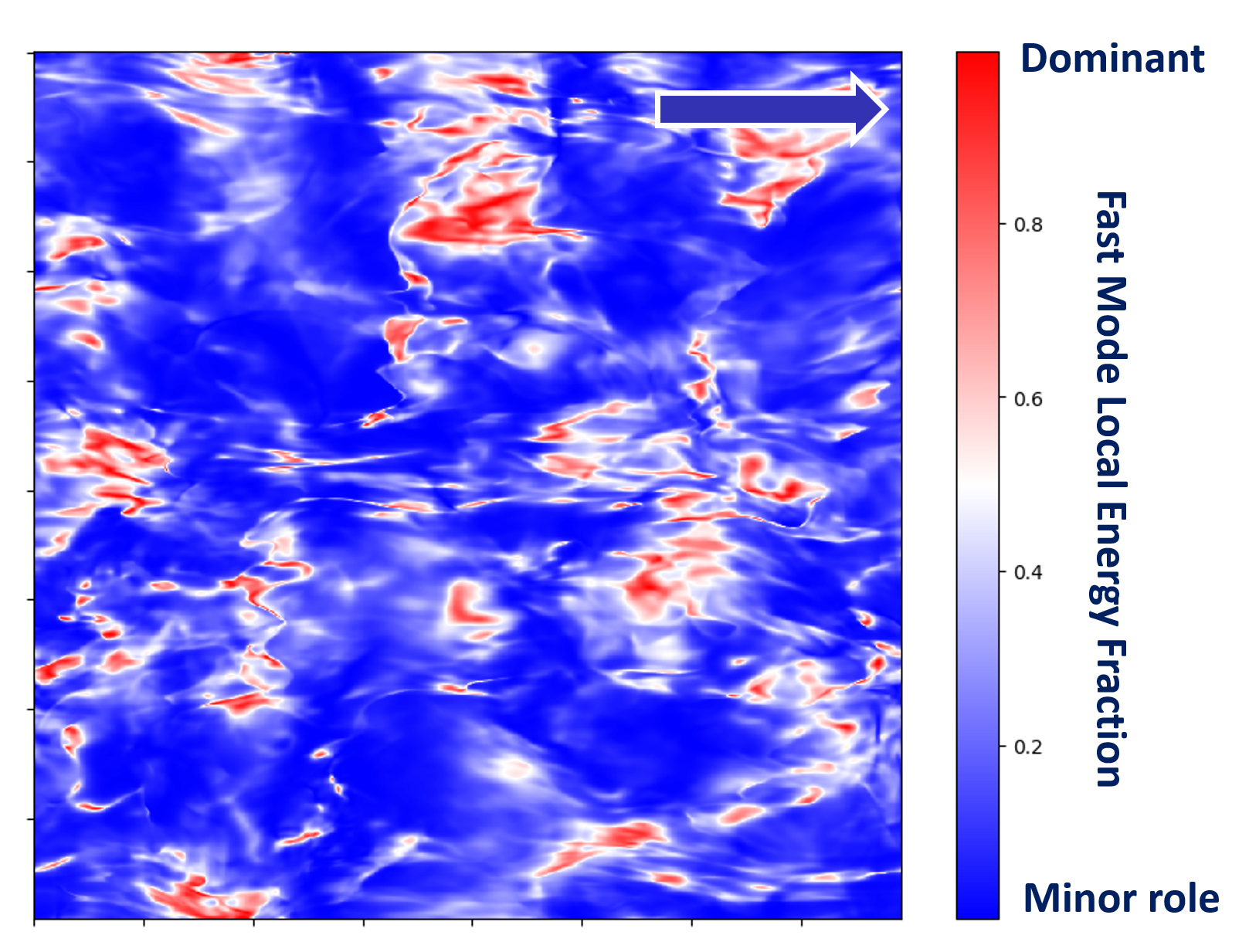}
\caption{illustration of the intermittency of fast mode in 2D slice. Maps of the fraction of fast mode energy fraction in each location from a slice map of 3D cubes. The region with red color represents fast mode is dominated in that sub-region, blue means playing the minor role and white means critical case. Color bar: Diverging color bar with 0.5 at white color, red when energy fraction approach to 1 and blue when energy fraction approach to 0. \\
Simulation used: H1S}
\end{figure}

If we talk about the local energy distribution of modes in 3D space, the situation is different in localized regions. Our simulations testify a very intemittent spatial energy distribution of the fast mode. For each cell in our simulation cubes, we get the total kinetic energy by adding up the kinetic energy of the Alfvénic modes $E_A$, the slow modes $E_s$ and the fast mode $E_f$ . We then divide the  energy of the individual mode to the total energy to study the relative distribution of turbulent modes in space. 
{\toreferee Mathematically, $E_{A,S,F} = \int_V \sum_{i}^{x,y,z}  v^2_{(A,S,F),i} dV$ and total energy $E_{total} = E_A + E_S + E_F$}.

In fact, we find that the energy of each mode is not distributed evenly across the space. strong concentration of fast modes is observed in  localized regions. Figure \ref{fig: Fast_Intermittency_2D} shows a visualization of this clustering effect from a slice of map from 3D cubes.  We define the regions with fast mode occupying more than 51\% of energy as fast mode dominated region. As shown in the figure, fast modes play a minor role in most of the volume. However, in the special localized regions, they would play the dominant role with up to 90\% of the energy are being in the form of the the fast modes. 

\begin{figure*}
\label{fig: Fast_Intermittency_3D}
\includegraphics[width=0.64\paperheight]{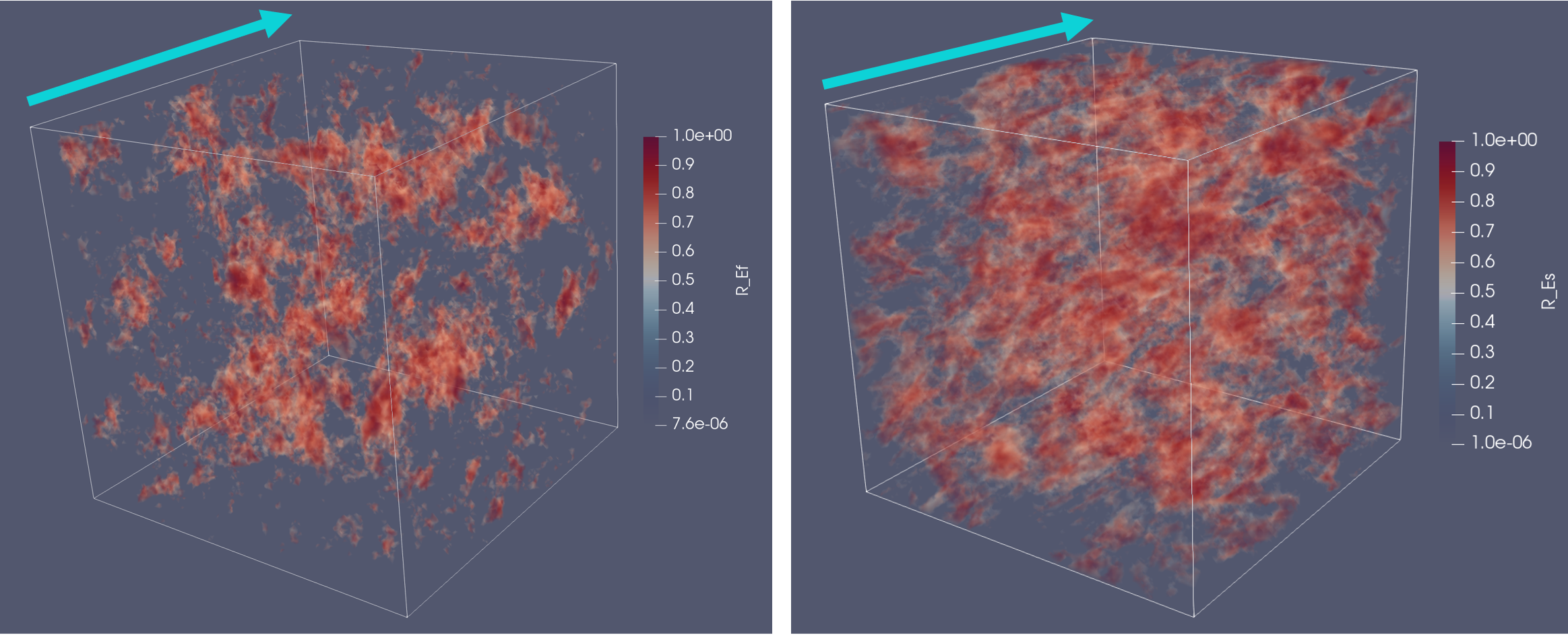}
\caption{3D Illustration of the intermittency of compressible mode in 3D using simulation H1S. Left: Fast mode Right: Slow mode. Maps of the fraction of mode energy in each location of 3D cubes.  Arrows indicate the mean direction of direction.} Regions coloured red represent sub-regions where the selected mode dominates, for instance, with fast mode energy fraction approaching 1 in dark red regions and 0 in transparent ones.
\end{figure*}

We notice that the fast mode dominated regions are not only clustering in the 2D slice but also continuous throughout the 3D space to form a 3D cluster. Figure \ref{fig: Fast_Intermittency_3D} visualizes this 3D property using the simulation data. The size of those regions occupy about 10\% of the total cube volume, but varies from simulation to simulation. We also searched for relationships between the fast mode-dominated regions and other MHD qualities, such as magnetic field strength and velocity amplitudes, but found no such relations. The location of those regions behave as a random variable with no relation to the local physical variables. 

\subsection{Intermittency of slow modes}

In addition, we noticed that the intermittency is also found in the slow modes. Same as fast modes, energy distribution of the slow modes is not evenly distributed in 3D space and clustering in the local region. About 30\% of the total cube volume is dominated by the slow modes. The right panel of Figure \ref{fig: Fast_Intermittency_3D} visualizes this 3D property of slow modes. In contract to fast modes, the slow mode shows stronger clustering effect and occupy more volume. However, most of them are concentrate in the component that parallel to the magnetic field. 

All in all, the energy of compressible modes is not evenly distributed across the space but concentrates in some local regions. The regions of fast mode dominance distort magnetic field tracing using the gradient technique. Indeed, one may expect that some of the local regions may be dominated by the fast mode region which, according to \cite{LY18a} can result in gradients that are orthogonal to the gradients arising from the dominant Alfv\'{e}nic modes. The slow modes are domination for the low $\beta$ is not expected to affect the magnetic field tracing.

\subsection{Intermittency of each modes throughout the evolution of simulation}
\begin{figure}
\label{fig: Mode_time_test}
\centering
\includegraphics[width=0.49\textwidth]{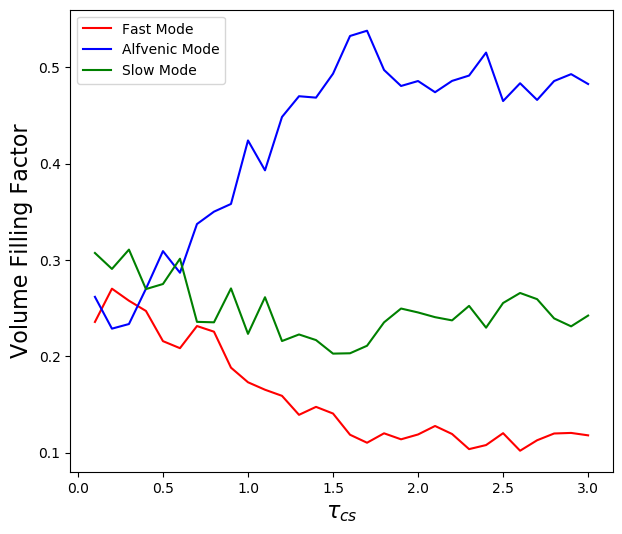}
\caption{The volume filling factor of dominated regions for each modes using simulation H0SS. }
\end{figure}

One may concern the properties of intermittency of each modes throughout the evolution of turbulence, We further study the change of intermittency using the simulation H0SS. For each snapshot ($\Delta t = 0.1 \tau_{cs}$), we compute the volume filling factor of dominated regions for each modes. We define the dominated regions as the single mode energy occupies more than half of the total energy in that unit volume. We notice that the sum of volume filling factor of dominated region of three modes will not be 1 as some of the unit volumes are not dominated by single modes. Figure \ref{fig: Mode_time_test} shows the result. One can see that the volume filling factor is about 0.3 for all modes at the beginning of the simulation at {\toreferee $t \approx 0.1 \tau_{cs}$}. The volume filling factor of Alfvenic mode then starts to grow steadily when another two modes drop. The system reaches a steady state at about {\toreferee $t \approx 1.7 \tau_{cs}$}. The volume filling factors are then stable with a mild fluctuation throughout the simulation until the end of simulation at {\toreferee $t \approx 3\tau_{cs}$}. The result at stable stage is consistent with our observation at previous subsection for all of the modes, which Alfvenic mode occupy half of cubes, about $25\%$ for slow modes and about $10\%$ for fast mode.

\section{Gradients for different MHD modes}
\label{sec:gradient modes}
\begin{figure*}[t]
\label{fig: Mode_Gradient}
\centering
\includegraphics[width=0.64\paperheight]{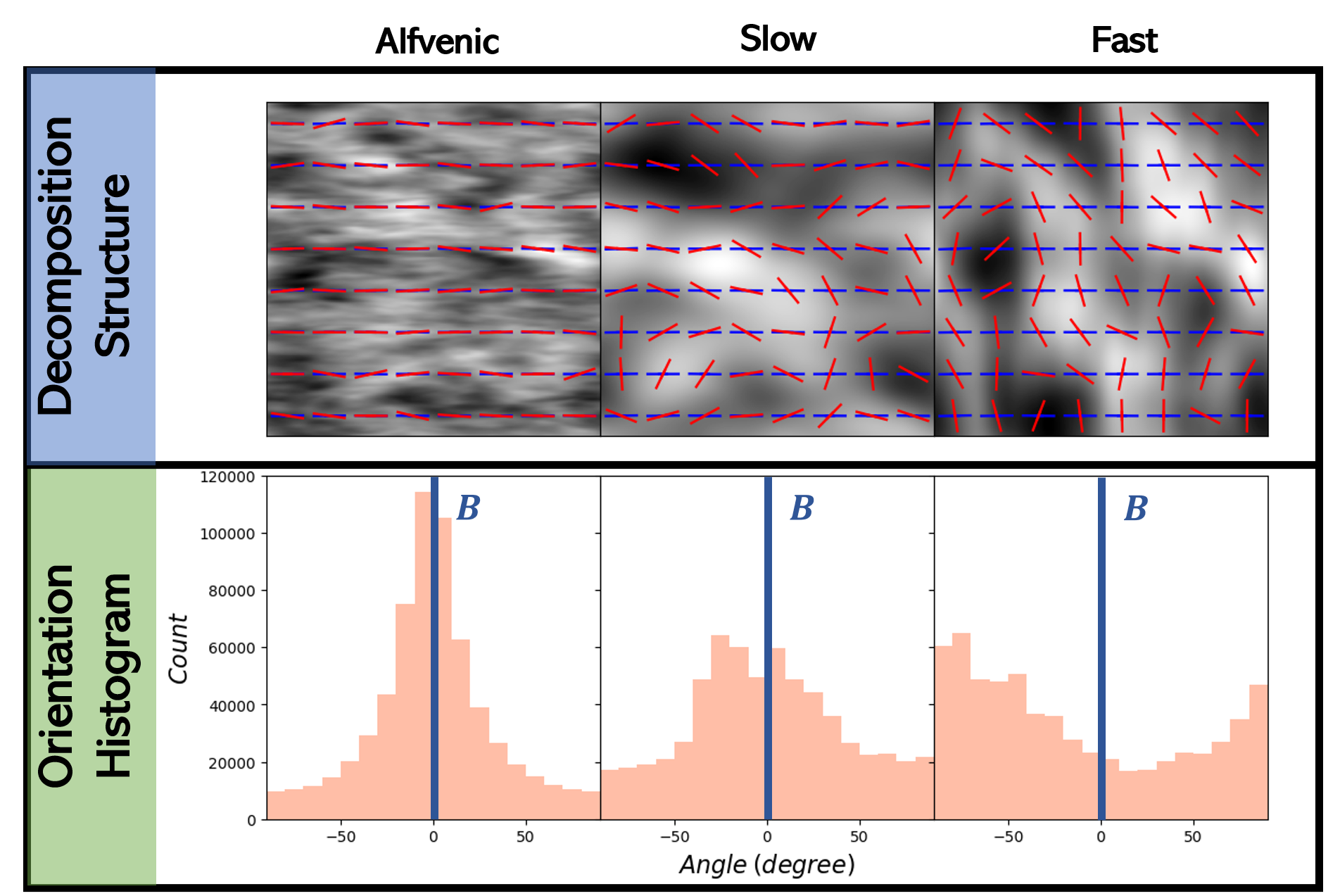}
\caption{Upper panel: The comparison between the Velocity Gradient and polarization from our simulation cubes used Figure \ref{eq:modes_computation}, the red and blue arrow indicate the local gradient (Marked in read) and magnetic field vector (Marked in blue). For the computation of local vectors, we employ the sub-block averaging method with the block size = $99^2$.\\
Bottom panel: The gradient orientation histogram for all the global gradient vectors in each modes. Global magnetic field orientation labelled in blue arrow. \\Simulation used: H1S.}
\end{figure*}
\subsection{Calculations of Gradient}

As a tracer of magnetic field, the gradient technique aim to trace the local alignment of the eddies with the local magnetic field. Below we calculate of gradients in the compressible mode, the slow and fast modes.

To calculate the gradient, we follow the procedure introduced in \cite{YL17}. We first compute the gradient field of the observable measures through the Sobel kernel. The whole gradient maps will then be divided to serval sub-blocks. The number of sub-blocks depends on the resolution of the map and the block size. Then we adopt the sub-block averaging method to probe the peak in gradient orientation distributions in the sub-blocks of the gradient map. The peak orientation of the gradient orientation distribution represents the direction of the projected local magnetic field and its dispersion would imprint the information of the local magnetization (see Lazarian et al. 2018). 

To quantify how good is the gradient field aligned with the magnetic field, we employ the alignment measure that is introduced in analogy with the grain alignment studies (see \cite{L07} ):
\begin{equation}
\label{eq:AM}
\begin{aligned}
AM = 2\left<\cos^2\theta_r \right> -1,
\end{aligned}
\end{equation}
with a range of $[-1,1]$ and $\theta_r$ represents the relative angle between gradient orientation and the orientation of magnetic field/polarization. Perfect alignments will give $AM=1$  whereas $AM=0$ for random alignment and $AM=-1$ for orthogonal alignment. This measure was first used for gradients in \cite{GL17}. Later it was adopted by other authors who study other statistics (see \cite{Soler19}).

\subsection{Properties of gradients for different MHD modes}

Previous studies explored the characteristics of gradient in different observations, like syncthrotron intensity maps \citep{LY18b} and spectroscopic channel maps \citep{LY18a} through the employment of mode decomposition method described in \cite{CL02,CL03}. Figure \ref{fig: Mode_Gradient} illustrates the features of gradient in the separated MHD modes from one of the cubes using the ZEUS simulation. It shows the results of the decomposed velocity cube,  projected along the x-axis. The upper panel demonstrates a comparison of the rotated gradient vectors and magnetic fields, which demonstrates the dependence of gradient and magnetic field in the local scale. For the bottom panel, we show the orientation histogram of rotated gradient vectors for each modes to visualize the global dependency. Consistent with the literature, the Alfvénic modes has high alignment between the gradient vectors and the magnetic field and shows very strong dependency for both local and global case. 

For slow modes, since they do not evolve on their own, but are sheared by Alfvén modes gradients also show the alignment between magnetic field and gradients. However, looking at the local alignment and orientation histogram in Figure \ref{fig: Mode_Gradient}, we see that the gradients arising from slow modes are less aligned compared to Alfv\'{e}nic modes. Slow modes are less aligned with the magnetic field locally and stronger signature of alignment in the global scale. However, its distribution is more disperse with respect to Alfv\'{e}n modes.

In contrast, gradient in fast modes have a totally different anisotropy. We see the local gradient vectors shows a significant perpendicular features in the local scale while orientation histogram also demonstrate a global tendency take perpendicular to magnetic field.

As a short summary, in terms of using gradients to trace magnetic fields, Alfvén and slow modes contribute to the expected alignment between the gradient and the magnetic field while fast modes play a disruptive role.

\subsection{Regions with perpendicular gradient alignment in observations}
\label{sec:intermittency}

We discussed the behaviour of gradient arising from different MHD modes in the last sub-section. In the real world case, the environment of MHD turbulence is composed of a mixture of the three basic modes. We know that the most important role for the MHD turbulence are Alfvén modes \citep{CL03}. As discussed before, Alfvén modes and slow modes together dominate MHD turbulence comprising more than 80\% of total energy. Therefore, one should expect gradient aligned with the local magnetic direction. However, gradient in observations can be affected by fast mode dominated regions.

\begin{figure}
\label{fig: Fliped region}
\includegraphics[width=0.49\textwidth]{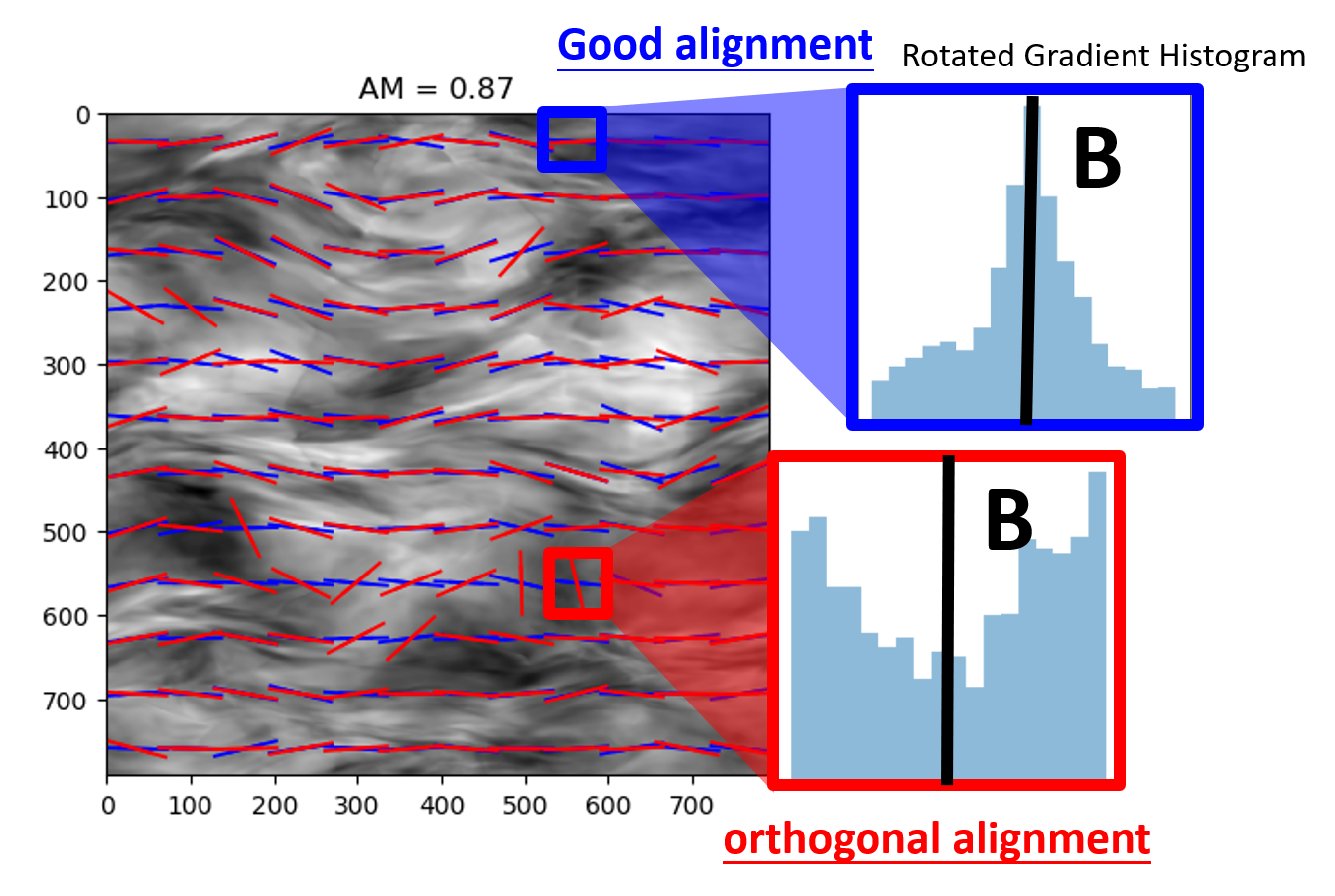}
\caption{illustration of the comparison between the good alignment region and the orthogonal alignment region in the 3D slice velocity map. The red and blue arrows represent the local magnetic field and gradient vectors. Blue and Red box shows the local orientation histogram of gradient vectors with local magnetic field labeled as black bar. Simulation used: H1S.}
\end{figure}

In Figure \ref{fig: Fliped region}, we demonstrated the alignment between gradient and magnetic field from a cross section map of 3D LoS velocity cubes. The map shows high alignment measurements with AM = 0.89, which indicates  gradient is highly aligned with the magnetic field in most of the regions. However, one can notice that gradient vectors are {\it perpendicular} to magnetic field in some of the local regions. This tendency is also shown by the orientation histogram in Figure \ref{fig: Fliped region}. The gradient orientation histogram shows the distribution of 90 degree rotated gradients perpendicular to the local mean field in that region.  We called it an {\it orthogonal} alignment and the perpendicular region as the Orthogonal Regions (ORs). This property have not been noticed before in the previous studies of gradient. In order to study magnetic field though gradient, it is very important to understand the underlying cause of the orthogonal region and the way to identify them.

\section{Fast modes and the orthogonal regions}
\label{sec:Intermittency}
\begin{figure*}[t]
\label{fig: FlippedRegionStudy}
\centering
\includegraphics[width=0.64\paperheight]{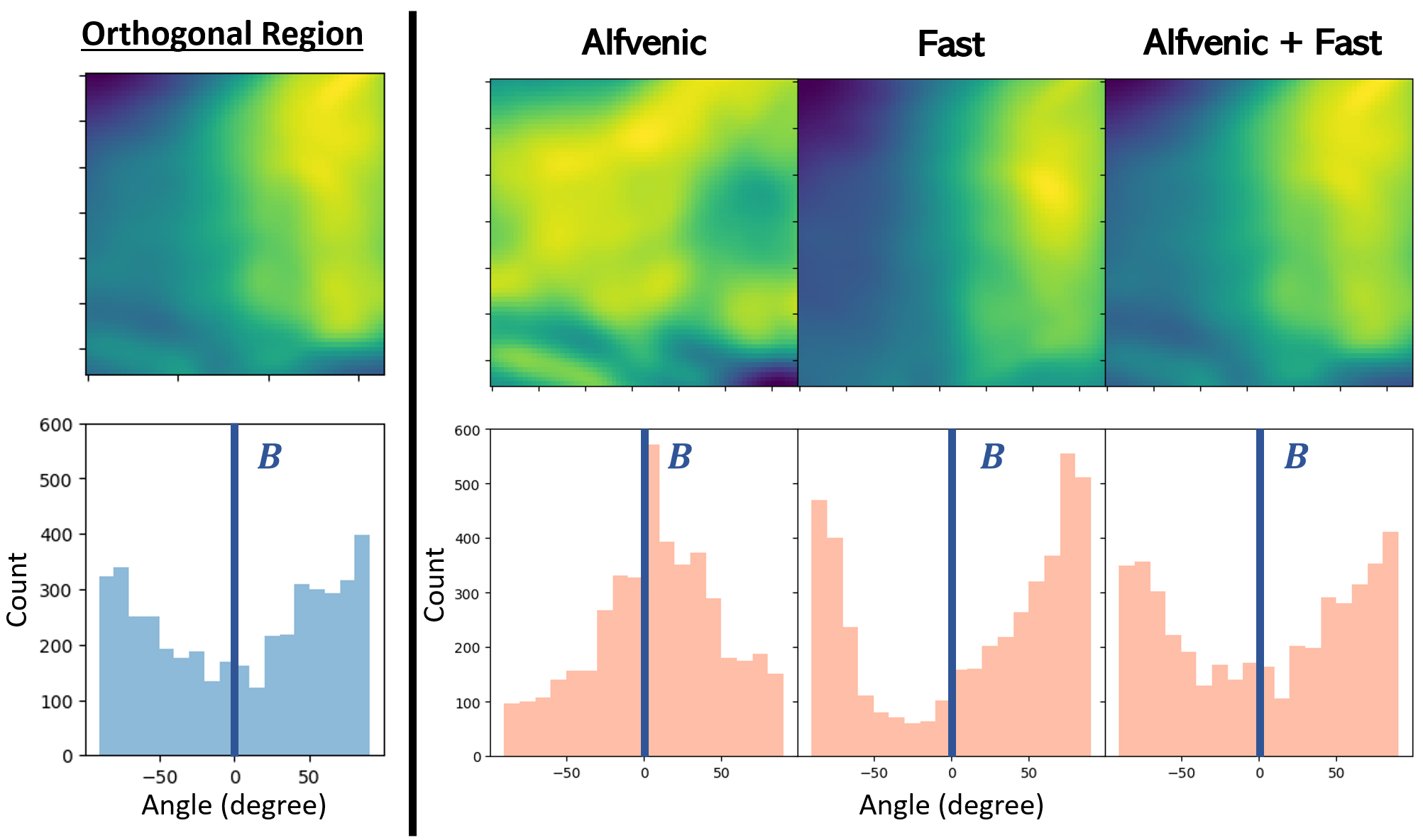}
\caption{Upper Left : LOS Velocity map of Orthogonal region before decomposition \\
Bottom Left : Gradient orientation histogram of Orthogonal region before decomposition \\
Upper Right : LOS Velocity map of Orthogonal region of Alfv\'{e}nic, Fast and superimposed modes\\
Bottom Right : Gradient orientation histogram of Orthogonal region of Alfv\'{e}nic, Fast and superimposed modes\\
Local magnetic field direction of the region marked as blue line with label B in each orientation histograms\\
Simulation used: H1S}
\end{figure*}

\subsection{Orthogonal regions in 3D space}

We further study the linkage between ORs and the intermittency of fast mode in this section.
Figure \ref{fig: FlippedRegionStudy} shows the decomposition of the orthogonal region and gradient orientation histogram we just showed in Figure \ref{fig: Fliped region}. We also add the pre-decomposed data in the left panel of the figure as a reference. As shown in Figure, we notice that the morphology of the orthogonal region have high similarity with its fast modes, a roundish structures located in the left, are different from  Alfv\'{e}nic modes, which have filamentary structure located in the bottom left corner of the figure. We have not shown the structure of slow mode of this orthogonal region but provide the corresponding discussion in the Appendix A. The local orientation histogram shows the correct anisotropy with respect to different modes but with lower dispersion compare to the pre-decomposed data, which means higher statistically significant anisotropy. The results of decomposition and histogram together indicate that the orthogonal region is a superposition of two MHD modes with fast modes being dominate. We further check the energy distribution of the energy of that region and find that about 90\% of the energy corresponds to the fast modes while the Alfv\'{e}nic modes take the remaining energy. To verify this hypothesis, we superimpose the Alfv\'{e}nic modes and fast modes of that region and compare the orientation histogram and structure to the pre-decomposed data. We put the superimposed result in the right panel for reference. The superposition of fast and Alfv\'{e}nic modes recovers well the pre-decomposed data.

This result, (see also section \ref{sec:intermittency}) provides a description of the ORs in 3D space. Due to the fast modes high intermittency, the clustering effect of the fast modes dominated in some localed region, which makes the anisotropy of velocity gradients change their directions by 90 degrees.

\subsection{Statistical analysis of observable measures, Reduced Centroid}

The situation of observable measures becomes complicated as a result of projecting of different MHD turbulence components. The projection mixes up the contribution from three modes along the line of sight.  The influence of fast mode being subdominant in observable measures for most of the region as Alfv\'{e}nic modes provide the major contribution along the line of sight.

We demonstrated the effect of different modes using reduced centroids \citep{YL18}. The reduced centroid $C_R$ provides the insight to the importance  the  between different modes as the project can be treated as an average velocity along the line of sight and its fluctuation of each mode is linearly separable, which is 

\begin{equation}
\label{eq:Reduced Centoid}
\begin{aligned}
\delta C_R \propto \rho_0 \delta < v_{LOS} > \\
=  \rho_0 (\delta <v_{A,LOS}> + \delta <v_{f,LOS}>) \\
= \delta C_{R,fast} +  \delta  C_{R,Alfven},  
\end{aligned}
\end{equation}
where $\delta$ is the spatial differentiation operator of arbitrary direction and $< .. >_{LOS}$ is the averaging operator along the line of sight. As we mentioned in section \ref{sec:Mode properties}, the fluctuation behaviour of Alfv\'{e}nic and Fast modes is bi-model, with Alfv\'{e}nic modes reaching its maximum fluctuation in the direction perpendicular to the local B-field lines and fast mode parallel to B-field lines. Also, slow mode can treated as the pseudo-Alfv\'{e}nic modes and follows the same properties.  So, if the amplitude of each mode is the crucial factor of deciding which mode takes a leading or dominating role and also the local anisotropic direction. In the language of gradient in reduced centroid map, the velocity fluctuation is equivalent to gradient amplitude($GA$). In a 2D observable measure like $C_R$, $GA$ accounts for the absolute fluctuation of both parallel and perpendicular B-field direction. We can use the the gradient amplitude as the indicator to study the mixture effect between the fluctuation of Alfv\'{e}nic and fast modes. For a more precise mathematical description, we define the gradient amplitude as the Euclidean norm of the gradient field in our computation procedure  (Reader may refer back to section \ref{sec:gradient modes} for the calculation of gradient.).

\begin{figure*}[t]
\label{fig: GA_statisitics}
\centering
\includegraphics[width=0.64\paperheight]{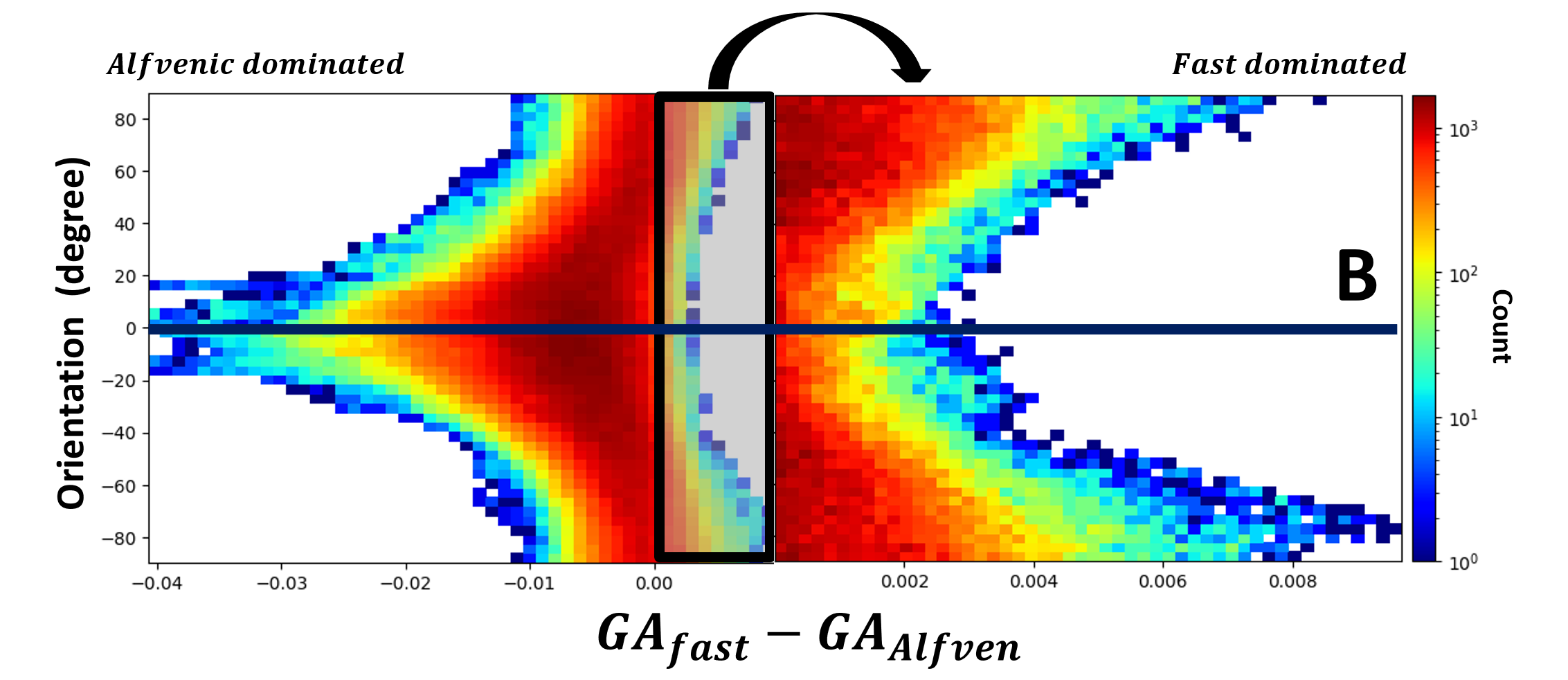}
\caption{2D histogram between the gradient amplitude difference and the gradient orientation in reduced centroid map for the simulation H1S. The color bar on the left demonstrated the bins number. Left plot showed the full 2D histogram and the right plot showed the magnified view of the 2D histogram of the fast mode dominated regime. The black line labelled the global magnetic field direction}
\end{figure*}

To imitate the behaviour of fluctuation in the local fast mode dominated regime, we perform a synthetic test for reduced centoid map with amplifying the fast mode energy. We amplify the fast mode velocity such that the kinetic energy ratio between Alfv\'{e}nic and fast modes is one to one in the global environment. To study the relation between the fluctuation of each mode and the gradient orientation, we perform a 2D histogram of gradient amplitude difference and gradient orientation to study this relation statistically. For each data point, we decomposed the contribution from fast and Alfv\'{e}nic modes and computed the difference of their gradient amplitude, then we computed their gradient orientation without the decomposition. Figure \ref{fig: GA_statisitics} showed the result using simulation H1S. 

In Figure \ref{fig: GA_statisitics}, We define gradient amplitude difference as $\Delta GA = GA_{fast} - GA_{Alfven}$. $\Delta GA$ characterises the mixture effect from modes in the 2D map, with a positive value indicating a fast mode fluctuation-dominated regime and a negative value representing Alfv\'{e}n mode fluctuation domination. As a reference, we put the direction of the global mean magnetic field in the figure. Figure \ref{fig: GA_statisitics} demonstrate a consistent picture with respect to properties of fast and Alfv\'{e}nic modes. For Alfv\'{e}nic modes dominated regime, a clear relation is found between the magnetic field and the gradient amplitude. With greater  amplitude difference in the negative side, meaning that Alfv\'{e}nic modes dominated regime, its distributions of gradient orientation is more concentrated to the global mean field direction. On the contrary, fast modes dominated regime in the positive shows a reverse relation, the distributions of gradient orientation is more concentrated to the direction perpendicular to the global mean field direction. For the region where $GA_{fast}\sim GA_{Alfven}$, since the anisotropy from one modes would being cancelled by another mode, no preferred direction should be found. As a result, when $\Delta GA$ approached 0, we saw a broader dispersion of values, with gradient orientation randomly distributed across all directions.

\subsection{Synchrotron intensity case}

Similar argument could also apply to other observable measures such as synchrotron intensity map $I_S$. As the physical environment of synchrotron is hot and diffuse, it is an sub-Alfv\'{e}nic and sub-sonic environment. Density fluctuation is moderate and mean field is much stronger compare to its fluctuation. We can then write the fluctuation of synchrotron intensity map $\delta I_S$ as:
\begin{equation}
\label{eq:Reduced Centoid}
\begin{aligned}
\delta I_S \propto \rho_0 \delta < (B_{0,POS} + \delta B_{POS})^2 > \\
=  2\rho_0<(B_{0,POS} + \delta B_{POS})(\delta B_{A+S,POS} + \delta B_{F,POS})> \\
\approx  2\rho_0B_{0,POS}(<\delta B_{A+S,POS}> + <\delta B_{F,POS}>) \\
\end{aligned}
\end{equation}
Where $B_{0,POS}$ is the mean field strength of the plane of sky(short write as POS),  $\delta B_{POS}$ is the fluctuation of the magnetic field which can be decomposed as Alfv\'{e}nic modes $\delta B_{A,POS}$, slow modes $\delta B_{S,POS}$ and fast modes $\delta B_{F,POS}$. We group the Alfv\'{e}nic and slow modes together since they share the same anisotropy. The expression comes with the same conclusion as the reduced centroid, which is that the anisotropy of the synchrotron intensity map depends on the amplitude of the fluctuations between the modes. Furthermore, as synchrotron intensity includes two components, the impact of fast mode would be impaired and gradient would be more aligned with the magnetic field. Studies of synchrotron intensity gradient showed a better accuracy of magnetic field tracing \citep{Letal17}.

To summarize, the anisotropy of observable measures depends on the comparison between the total fluctuations of each of the mode along the line of the sights. 

\begin{figure}
\label{fig: AF}
\includegraphics[width=0.49\textwidth]{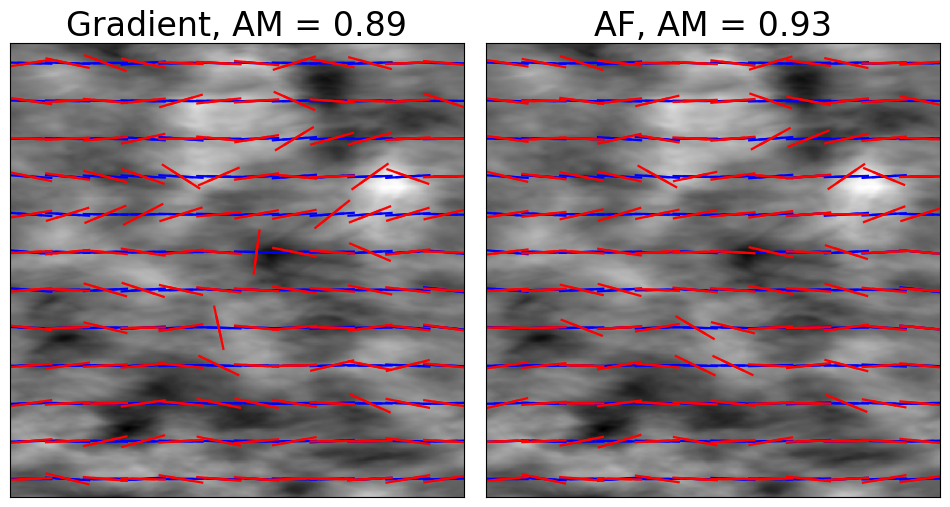}
\caption{Centroid gradients(left panel) and Centroid gradients with AF (right panel). Block size of both maps are $66^2$. AM of each method is labelled at the top. Threshold of filter: GA $\sim$ 0.01\\
Simulation used: H1S}
\end{figure}
\section{Reducing the impact of fast modes on gradient technique}
\label{sec: tech}
\subsection{Amplitude Filtering (AF) technique}
In the last sub-section, we observed that the net gradient amplitude difference between the Alfv\'{e}n modes and fast modes showed in Figure \ref{fig: GA_statisitics}. Even though we assumed the equal global energy between two modes, we noticed that the amplitude of Alfv\'{e}n modes are still few times larger than the fast modes. As a comparison showed in the figure, the maximum amplitude of the fast mode dominated data points are only $\sim 0.01$ while Alfv\'{e}nic modes are $\sim 0.04$. This indicates that the pixels with large value of gradient amplitude in centroid or synchrotron intensity map would essentially dominated by the fluctuation of the Alfv\'{e}n modes. So, we can reduce the impact of fast mode by utilizing the effect of dominating Alfv\'{e}n fluctuation at large gradient amplitude. 

To remove the pixels dominated by the fast mode, one can remove pixels with low value of gradient amplitude. Those pixels contain the equal mixture of the fluctuation between the Alfv\'{e}n and fast modes, which resulting in a random orientation distribution with no peak direction. The remaining pixels would then mainly be coming from the Alfv\'{e}n modes and having a clear peak direction, which is the local mean direction of magnetic field or polarization.

In Figure \ref{fig: AF}, we compare the magnetic field tracing with Centroid gradients using raw data and using the AF approach. For the latter, the filter suppressing regions of low amplitude gradients was applied. We observe an increase of the AM from 0.89 to 0.93. 

We also noticed that this method would be incapable of removing the contribution of fast modes at the supersonic regime. As shocks are created at the supersonic regime, the density effect would plays an important role in the shock region. The mass would accumulate at the shock front as the shock push across the space. Taking an example of the radiative shock at sub-Alfv\'{e}nic environment, one that is usually true in molecular cloud, the density behind and after the shock would have a difference of ratio of $M^2_s$. This would create a large gradient amplitude and having an anisotropy that is perpendicular to the mean field as shock  favours pushing in a direction of parallel to the mean field.

\subsection{Gradients of Gradient Amplitude (GGA)}

Beside of the gradient amplitude filtering method, can also use gradients of gradient amplitude (hereafter GGA). The study of gradient amplitudes was first done in \cite{YL20GA}. There the gradient amplitudes were used to obtain the sonic Mach number of turbulence, i.e.
$M_s=V_L/V_c$, where $V_L$ is the turbulence injection velocity and $V_c$ is the sound velocity. Here we suggest a new way of using the gradient amplitudes, i.e. to calculate the gradients of the gradient amplitudes and use the GGA to trace magnetic field. 

\begin{figure*}[t]
\label{fig: GGA}
\centering
\includegraphics[width=0.64\paperheight]{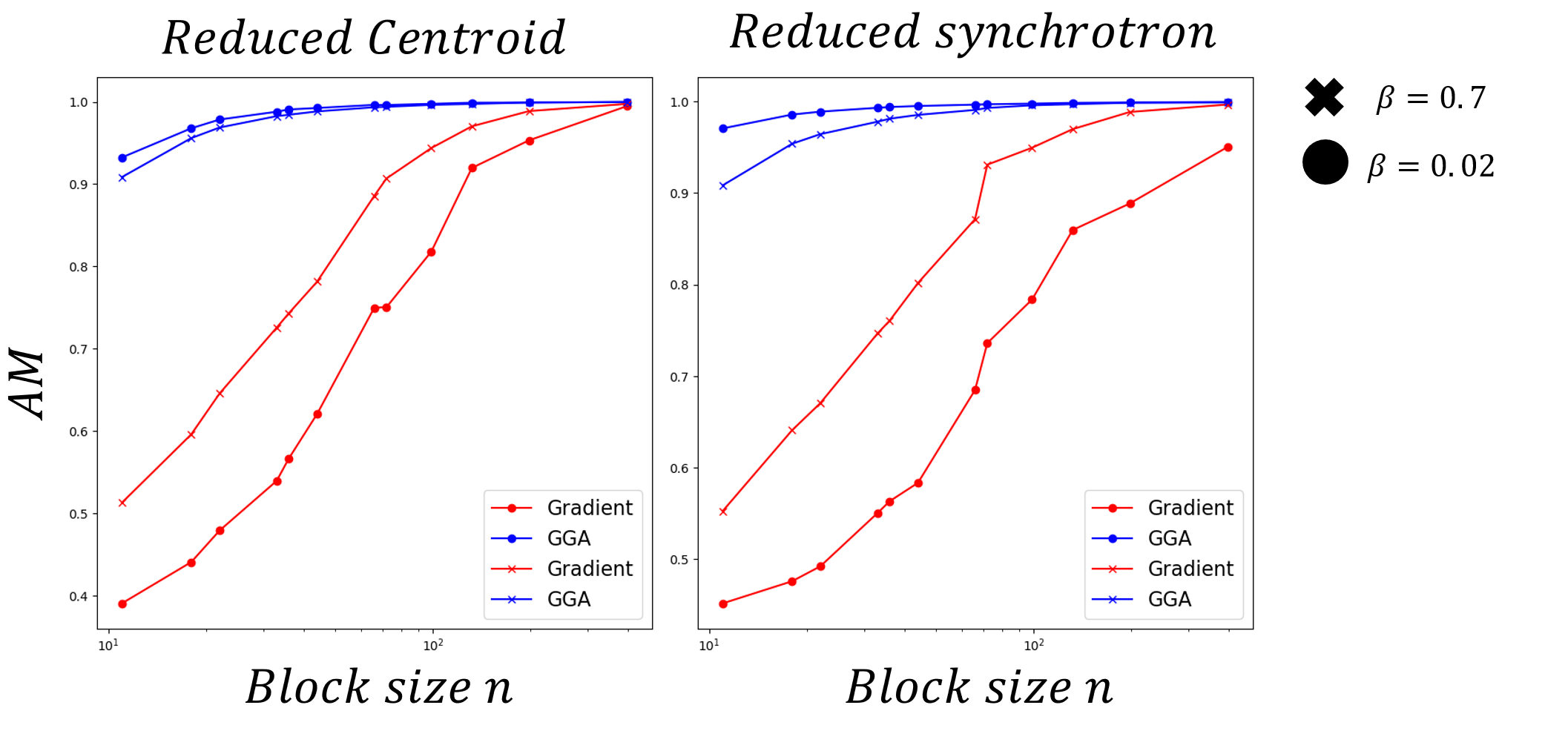}
\caption{Left: The variation of AM to block size of gradient and GGA for two synthetic reduced centroid map. Right: The variation of AM to block size of gradient and GGA for two synthetic reduced synchrotron intensity map map. The data point with dot marker indicates the simulation of $\beta = 0.02$ and cross marker for $\beta = 0.7$.\\
block size covered : $[11,18,22,33,36,44,66,72,99,132,198,396]$}
\end{figure*}

Figure \ref{fig: GGA} shows the comparison of the SIGs maps and GGA maps obtained with synchrotron intensities. To obtain the latter map, instead of using the raw observation maps as a input, we input the gradient amplitude map for the calculation and getting the gradient direction from the gradient field of gradient amplitude. The comparison of the two maps shows that GGA mitigates the influence of the fast modes and shows a superior performance  in magnetic field tracing compared to the SIGs. For the GGA we used the standard procedure of sub-block averaging \cite{YL17a}. 

\begin{figure}
\label{fig: SIG_GGA}
\includegraphics[width=0.49\textwidth]{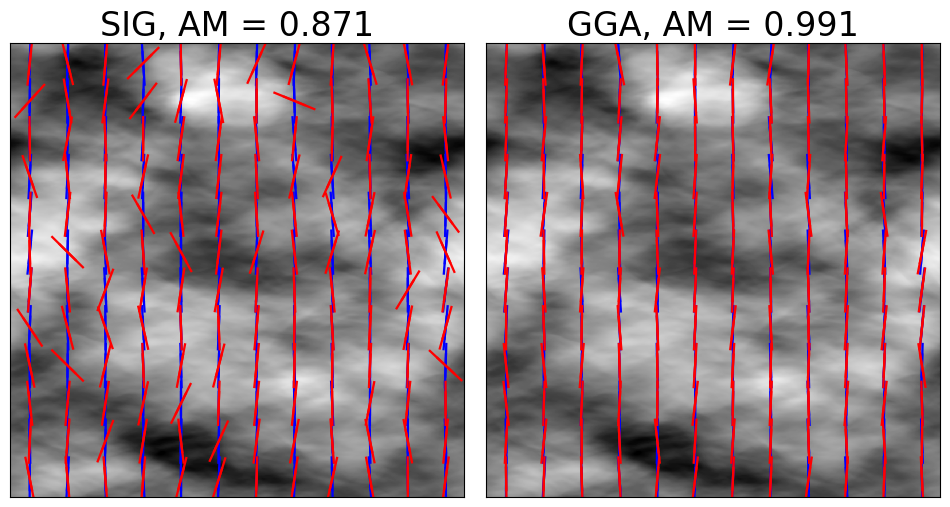}
\caption{Synchrotron Intensity gradients(SIG, left panel) and GGA(right panel). Block size of both maps are $66^2$. AM of each method is labelled at the top.\\
Simulation used: H1S}
\end{figure}

Applying GGA to velocity centroids, we also found that for reduced centroid, GGA technique demonstrate a better performance of tracing magnetic field. For the comparison between two techniques, we compute their alignment measure of different block sizes. Figure \ref{fig: GGA} showed the result. We see for both of the techniques, the AM increase with the block size. However, GGA shows a earlier saturation of smaller block size with the size of $\sim 66^2$ pixels compare to the larger size of $\sim 200^2$ for gradient technique. For both of the synthetic maps, the GGA performs better than gradient with the performance gap  of $\Delta AM \sim 0.4$ at the beginning of small block size. The performance gap then decreases gradually with the increase of block size.

Nonetheless, we also notice that the new GGA technique is more sensitive to noise, which is an important factor in handling observational data. The performance of GGA drops significantly with the increase of the noise. This limits the use of GGA to the observational data with low S/N ratio. We expect that the adoption of the smoothing filter , for instance, smoothing with Gaussian kernel, could restore the performance of GGA. We would studied the new technique exhaustively in the future study.  

\subsection{Using Velocity Channel Gradients (VChGs) and Combining GGA approaches}

As \cite{LY18a} showed, the decomposed channel map of the three modes also shared the same anisotropy properties with centroid and synchrotron intensity map. Unlike other observable measures imprinting the 3D information through projection, spectroscopic channel maps imprint the 3D information through reallocating the 3D information from real space to the LOS velocity space. The anisotropy property of channel map depends on the channel width chosen. The contribution of each mode in channel map cannot be separated linearly even when constant density is assumed. 

\begin{figure}
\label{fig: Ch_Cen}
\includegraphics[width=0.49\textwidth]{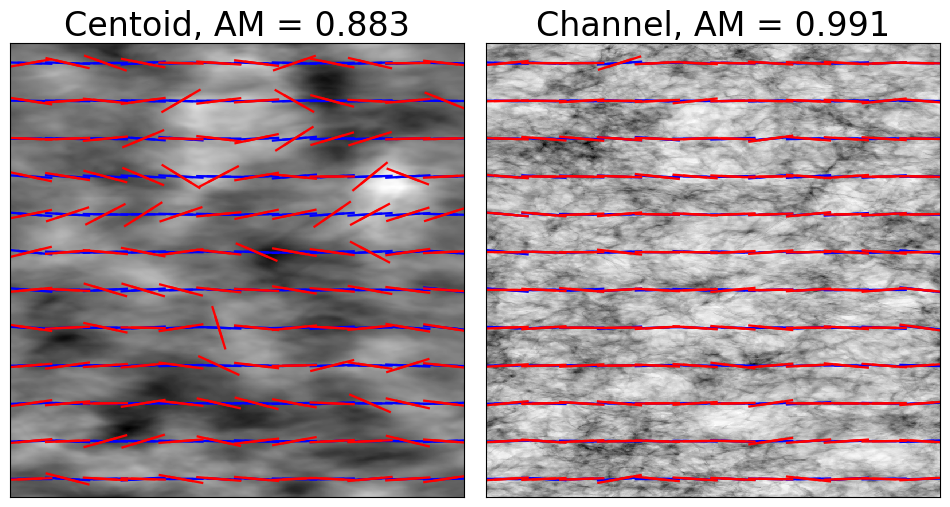}
\caption{Velocity centroid gradients( left panel) and Velocity channel gradient(right panel) using simulation H1S. Block size of both maps are $66^2$. AM of each method is labelled at the top. Width of channel $\Delta v$ used: 0.1 $\sqrt{ <\delta v^2_R>}$, where $\sqrt{<\delta v^2_R>}$ is the square root of the  turbulent velocity dispersion.}
\end{figure}

Demonstrated by \cite{LY18a}, gradient of velocity channel map carries the information of turbulent and its anisotropy. The thickness of the channel decides the information we trace, which are  turbulent velocities in thin channel and turbulent densities in thick channel. For thin channel maps, its alignment shows a better tracing performance of magnetic field direction than other techniques, such as correlation functions anisotropy analysis, velocity centroid gradient and reduced centroid gradient in both sub-sonic and super-sonic turbulence environment. In our study, we also notice that the impact of fast modes is being suppressed in the channel gradient using thin channel. In Figure \ref{fig: Ch_Cen}, we compare the magnetic field tracing with raw Centroid gradient (without using AF technique) and thin channel gradient. We observed that the increase of AM from $\sim 0.88$ to $\sim 0.99$. Also, the local locations that orthogonal gradients are happened in centroid map are also being suppressed in the thin channel gradient. 

\begin{figure}
\label{fig: PPV_GGA}
\includegraphics[width=0.49\textwidth]{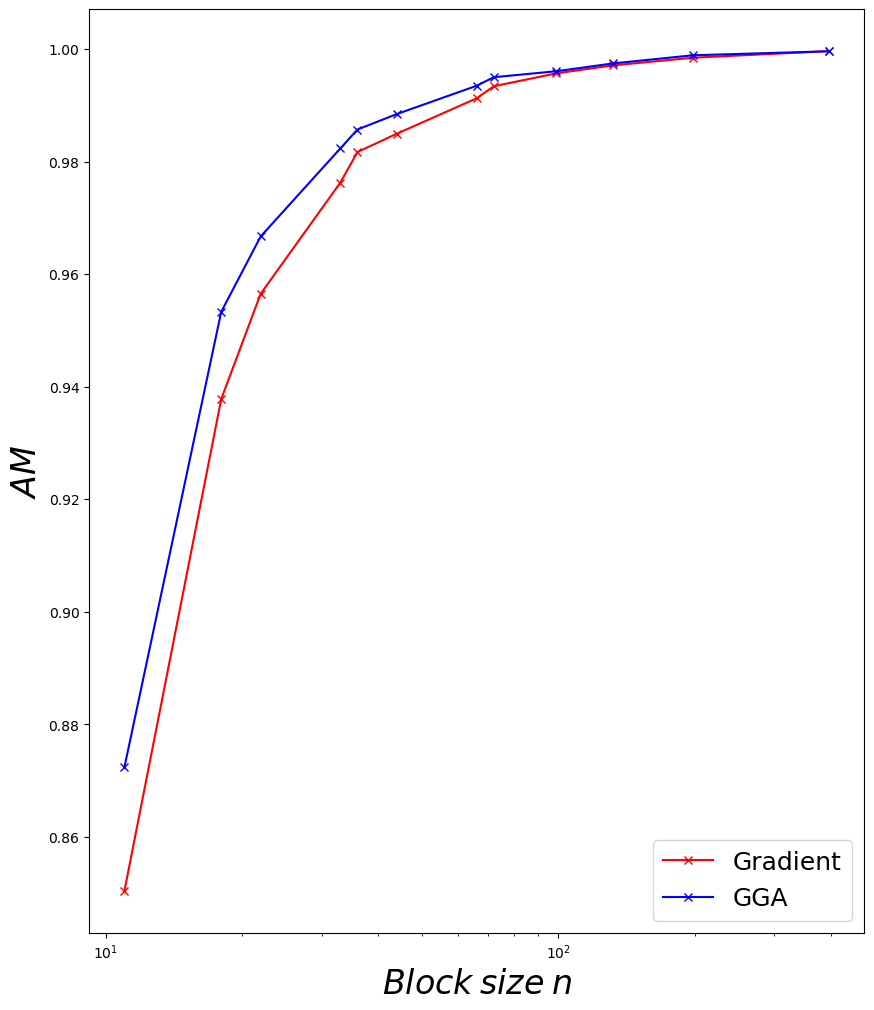}
\caption{The variation of AM to block size of channel gradient and GGA using simulation H0S.\\
block size covered :[11,18,22,33,36,44,66,72,99,132,198,396]}
\end{figure}

We also combined the GGA technique with channel maps for the further suppression of the fast mode. Compare to the raw channel gradients, channel gradients combined with GGA demonstrate a better performance of tracing magnetic field. However, the improvement of applying GGA to channel is insignificant in contract to Synchrotron intensity or Centroid. In Figure \ref{fig: PPV_GGA}, we show the result of the improvement of combining channel and GGA together with various block size using the simulation H0s. A slight improvement of $\Delta AM \sim 0.03$ is observed at the smaller block size and performance gap 
narrows down with the increasing block size. The performance improvement become negligible when the block size larger than $44^2$. The result indicates that the hybrid approach is suitable for the small block size.

\section{Discussion}
\label{sec:discussion}

\subsection{Intermittency of fast modes and its importance}

The properties of fast modes were studied in a number of earlier papers \citep{CL02,CL03,KL10}. In this paper we identified an important effect corresponding to the fast modes, namely, their intermittency in low $\beta$ turbulence. The consequences of the discovered property of fast modes should be explored further.

Fast mode plays a vital role in different astrophysical process, especially for cosmic ray (hereafter CR) scattering and acceleration. As turbulent energy is injected at large scales, fast mode is the major source of CR scattering in the interstellar and intracluster mediums (see \cite{YL02,YL04}). The earlier CR scattering treatment typically assumes a isotropic fast mode scattering and homogeneous distribution of fast mode energy (see \cite{YL08}). The location of CR scattering usually include the Galactic Halo and warm ionized medium, which is low-$\beta$ and sub-sonic region. The numerical result in this study would indicate an implication of physical picture current CR scattering model, the localized CR scattering. The CR scattering may happened in some localized region where fast mode energy is clustered.

\subsection{Relation to earlier studies}
We noticed that earlier studies has also found the intermittency effect of MHD compressible mode in the sub-sonic environment \citep{LFW2016,Park2019,MY20}. They have explored the spatial distributions of fast modes through various statistics. {\toreferee In  this paper,  we  further  explore  the  relationship  between  magnetic  field  direction, gradient orientation distribution  and  spatial  distributions  of  fast  modes.} We also study how the intermittency effect influence the anisotropy of observational measures such as centroid and synchrotron intensity.

For the gradient technique, it recently introduced a new way of tracing magnetic field. Its foundation is the theory of MHD turbulence, in particular, the properties of Alfv\'{e}nic modes. 

Velocity gradients as well as their counterparts, e.g. synchrotron intensity and synchrotron polarization gradients, were showing the good alignment with the underlying magnetic field, but the alignment was not perfect. The fast modes were suspected to cause the deviations, but it was surprising to see occasional significant deviations of gradients along particular directions. Indeed, in the simulations which were employed, the fast modes were subdominant compared to Alfv\'{e}n modes. 

In this paper through a careful analysis of the low-$\beta$ turbulence simulations we identified the reason for this behavior is that fast modes are very intermittent. Therefore, in spite of general subdominance, the fast modes can dominate the signal along given lines of shight.

\subsection{ Comparison of orthogonal alignment between Gravitating systems, Shock, and Fast Mode dominated system}
The gravitational-dominated system, shock region and fast mode dominated system all share the same gradient signature, an orthogonal gradient distribution in the gradient orientation histogram.

\cite{Hu2020,YL17b}  studied the feature velocity gradient in the presence of gravity. When the gravity is absent, consider the GS95 picture and we would see that the maximum change of velocity gradient is in the direction perpendicular to the local magnetic field. In the case of strong self-gravity, gravitational force is expected to modify the properties of flows in the vicinity of the centers of the gravitational collapse. Mass flows in the gravitational center and cause the negative divergence of velocity. This gravitational pull produces the most significant acceleration of the plasma in the direction parallel to the magnetic field, and the velocity gradients are parallel to the magnetic field. As a result, velocity field deviates from the normal magnetized turbulent fluid. The orthogonal gradient happens in the gravitational center. This is often closely related to the highly column density as mass is accumulated in the dense region. Additionally, \cite{Crutcher2010} showed this is a power-law relationship between the value of LOS component of magnetic field strength inferred from the Zeeman splitting and the column density. So, orthogonal alignment , density and magnetic field strength clasped tightly together in the gravitational dominated systems.

In addition, orthogonal alignment in fast mode dominated region is not necessarily related to the density or magnetic field strength. It depends on the local energy ratio in the observable measures. Component effect and the angle between mean field and LOS play a vital role here. Observable measures contain more than one component could include more weight of slow and Alfv\'{e}nic modes and balance the clustering effect of fast mode. We would expect synchrotron intensity experience a slight influence and a stronger influence for centoid.  Although the nature of intermittency provide a clustering effect throughout the 3D space, this effect is not noticeable and smooth out during the projection as it is not related to the density and magnetic field strength. 

On the other hand, study found the existence of parallel velocity gradients in ideal, compressible MHD simulation \citep{Beattie2020B}. As showed in section 4 of \cite{Beattie2020B}, the authors demonstrated that a large velocity gradient appears in the shock region and its gradient direction along the mean magnetic field. As a result, a orthogonal gradient could be found in those regions. We noticed that this phenomenon is closely related to the turbulent system that is both Sub-Alfvenic and supersonic. The location of the parallel gradient usually appears in the shock regions and its adjacent. Also, it is closely related to the density value as shock accumulated the mass. These two properties differentiate the orthogonal gradient that happened between the shock region and fast mode dominated system. As mentioned in the previous paragraph, we have not found a relationship between the density and fast mode dominated region but parallel gradient happens in shock region usually associated with high density and large velocity gradient amplitude. Those regions usually can be signified as a density structure with thin and long filamentary shape, which makes it easier to be identified compared to the fast mode dominated system (See \cite{YL17a,LY18a,YL20GA} for the identification of the shock in the high sonic Mach number regime).  

\section{summary}
\label{sec:summary}
This paper further study the properties of fast modes and its impact in both 2D and 3D space using the low-$\beta$ MHD simulations through the numerical method introduced by \cite{CL03}. Our main discovery are :\\
1. While the fast modes in low-$\beta$ MHD turbulence occupied the least energy in comparison to other modes, but the fast modes are very intermittent and dominated in localized sub-regions.\\
2. In those localized regions, the physical properties would change, for instance, the velocity anisotropy could be perpendicular to local magnetic field.\\
3. We also studied the impact of fast mode for typical observable measures such as centroid and synchrotron intensity maps. Even though the Alfv\'{e}n modes dominated along the line of sight, we found that the mean fluctuation caused by fast modes could makes a reflectable influence to the 2D observe measures.  For the local region that its fluctuation dominated by the fast mode, the anisotropy would be perpendicular to plane of sky magnetic field and therefore resulting in the orthogonal gradient.\\
4. Based on the development of VGT, we further developed the Amplitude Filtering (AF) technique and Gradients of Gradient Amplitude (GGA) technique to suppress the impact of slow mode. We applied the new techniques to synthetic centoid and synchrotron map and showed that they could reduce the impact of fast mode and therefore improve the tracing power of gradient technique.\\
5. We compared the orthogonal gradient happened in the collapsing self-gravitating region and fast mode dominated region. The difference between two are distinguishable as gravitational centers are often closely related to the high column density, strong magnetic field strength and high gradient amplitude.\\
6. We discussed that the intermittency of fast modes could be very important for many astrophysical applications, e.g. cosmic ray propagation and acceleration.

\appendix

\section{Component effect of MHD modes}
Our paper introduces the spatial distribution of fast mode energy is not uniformly distributed but with clustering effect. In this section, we would also discuss the energy distribution of each mode across different components as their features of prorogation deciding their energy distribution. 

Taking Alfv\'{e}nic wave as a starting point, in simple physical picture, Alfv\'{e}nic wave is transverse wave propagating along the magnetic field lines. As a result, fluctuation only allowed in the direction perpendicular to the local magnetic field lines. In other words, energy would only distributed in the $k_{\bot}$ direction. This feature would preserve in the turbulent environment as the structure function shows the fluctuation of $k_{\bot}$ direction decay in the 5/3 power law in contract to the $k_{\parallel}$ direction.  

For other two compressible modes which are fast and slow modes, the situation becomes more complicated since because of their compressible nature. Here, we would try to discuss their feature in the low-$\beta$ regime. we can start by looking at the unit vector form presentation of slow modes in \cite{CL03}: 

\begin{equation}
\label{eq:slow_mode_presentation}
\begin{aligned}
\hat{\xi}_s \propto \Big(1-\frac{\beta}{2}-\sqrt{D}\Big)k_{\bot}\hat{k}_{\bot}+\Big(-1-\frac{\beta}{2}+\sqrt{D}\Big)k_{\parallel}\hat{k}_{\parallel}
\end{aligned}
\end{equation}

By considering $\beta \ll 1$ and taking the Taylor expansion, the expansion can be simplified as;

\begin{equation}
\label{eq:slow_mode_presentation}
\begin{aligned}
\hat{\xi}_{s,\beta \ll 1} 	\propto \frac{\beta}{2}\Big( 2 cos^2\theta - \frac{\beta}{2}\Big)k_{\bot}\hat{k}_{\bot}   \\   
+ \Big(-2 + \frac{\beta}{2}\Big( 2 cos^2\theta - \frac{\beta}{2}\Big) \Big)k_{\parallel}\hat{k}_{\parallel}
\end{aligned}
\end{equation}

We would see slow modes depends on the $\beta$ and angle between the LOS and the mean field lines. Another point to make in here is that, it is apparently the $k_{\parallel}$ component dominate over the $k_{\bot}$ direction because of the $\beta /2 $ factor. Comparing the $k_{\parallel}$ and $k_{\bot}$ component, the most important consequence of slow mode would like the Alfv\'{e}nic modes but reverse in the low $\beta$ environment, almost no slow mode energy distributed to the direction that perpendicular to the magnetic field lines. 

On the other hand, the ratio between $k_{\parallel}$ and $k_{\bot}$ component of fast mode plays minor role here. The ratio does not change much and close to unity. It indicates that energy of fast mode is distributed uniformed for all direction. This match the feature of an acoustic-type cascade for fast modes as its prorogation is like sound wave, marginally concerned about the magnetic field direction.

As observational variables may only contain one or two vector components, this component effect makes a huge impact to the observational variables. The typical example would be the centroid. This could be showed by the reduced centroid map which contribution of velocity could be decomposed as a superposition of different modes:
\begin{equation}
\label{eq:Conponents_Centroid}
\begin{aligned}
C({\bf R}) = \rho_0 \int v_{LOS}({\bf R},s) ds \\
          = \rho_0 \int v_{A,LOS} + v_{S,LOS} + v_{F,LOS}\:ds \\
          = C_{A}({\bf R}) + C_{S}({\bf R}) + C_{F}({\bf R}),
\end{aligned}
\end{equation}
where $C_{A},C_{S},C_{F}$ are the separated contribution from each modes and $\rho_0$ is the constant density. As we just shown, the angle $\theta$ between the mean field vector and line of sight deiced the contribution of each modes, especially for slow and Alfv\'{e}nic modes. For $\theta \sim \pi/2$, Alfv\'{e}nic and fast modes contribute most of the anisotropy while slow mode plays an minor role on centroid. In this case, the fast modes play a more important role than slow modes. This effect would become moderate on the synchrotron intensity map since they contain two components, which average out the contribution of each modes.

\end{document}